\documentclass[a4,11pt,reqno]{amsart}%
\usepackage[english]{babel}
\usepackage[latin1]{inputenc}
\usepackage{graphics}
\usepackage{ulem}
\usepackage{hhline}
\usepackage{dsfont}
\usepackage{mathrsfs}
\usepackage{fancyhdr}
\usepackage{amsmath,amssymb}
\usepackage{rotating}
\usepackage[latin1]{inputenc}
\usepackage[T1]{fontenc}
\usepackage{fancybox}
\usepackage{color}
\usepackage{colortbl}
\usepackage{setspace}
\usepackage{enumerate}
\usepackage[nice]{nicefrac}
\usepackage{amsthm}
\usepackage{multicol}
\usepackage{pifont}
\usepackage[french]{minitoc}
\usepackage{latexsym,amsfonts}
\usepackage{amsthm}
\usepackage{varioref}
\usepackage{textcomp}
\usepackage{lmodern}
\usepackage{mathpazo}
\usepackage{euscript}
\usepackage[pdftex]{hyperref}
\numberwithin{equation}{section}
\makeatletter\@addtoreset{equation}{section}

\DeclareMathSymbol{\subsetneqq}{\mathbin}{AMSb}{36}

\begin{document}

\title[Probability distributions attached to generalized Bergman spaces]{Probability distributions attached to generalized Bergman spaces on
the Poincar\'{e} disk}
\author[N. Askour $\&$ Z. Mouayn]{\bf N. Askour and Z. Mouayn}
\address{{\bf N. Askour and Z. Mouayn} \newline {\small Department of Mathematics, Faculty of Sciences and Technics (M'Ghila),} \newline
{\small Sultan Moulay Slimane University, BP 523, B\'{e}ni Mellal, Morocco }}
 \email{askour@fstbm.ac.ma} \email{mouayn@fstbm.ac.ma}
\maketitle

\begin{abstract}
A family of probability distributions attached to a class of generalized
weighted Bergman spaces on the Poincar\'{e} disk are introduced by
constructing a kind of generalized coherent states. Their main statistical
parameters are obtained explicitly. As application, \ photon number
statistics related to coherent states under consideration are discussed.
\end{abstract}

\section{Introduction}

\noindent The \textit{negative binomial states} are the field states that
are superposition of the number states with appropriately chosen
coefficients \cite{1}. Precisely, these states are labeled by
points $z$ of the complex unit disk $\mathbb{D}:=\left\{ z\in \mathbb{C}; \, |z| <1\right\} ,$ and are of the form
\begin{equation}
\mid z ;2\beta >=(1-|z|^{2}) ^{\beta
}\sum\limits_{k=0}^{+\infty }\left( \frac{\Gamma (2\beta+k) }{%
\Gamma (2\beta) k!}\right) ^{\frac{1}{2}}z^{k}\mid k>,
\label{1.1}
\end{equation}
where $2\beta >1$ is a fixed parameter and $\mid k>$ are number states.

The probability of finding $k$ photons in the state \eqref{1.1} is
given by the squared modulus of the projection of $\mid z$ $;2\beta >$
onto the state $\mid k>$ as
\begin{equation}
\left| <k\mid z  ; 2\beta >\right| ^{2}=(|z|^{2}) ^{k}(1-|z|^{2}) ^{2\beta }\frac{\Gamma
(2\beta+k) }{\Gamma (2\beta) k!},k=0,1,2,\cdots
\label{1.2}
\end{equation}
The latter is recognized as the \textit{negative binomial distribution} $%
\mathcal{NB}\left( |z|^{2},2\beta \right) $ with $\left|
z\right| <1$ and $2\beta >1$ as parameters \cite{2}. Furthermore,
the probability distribution \eqref{1.2} has a positive Mandel
parameter and thereby the negative binomial states obey super-Poissonian
statistics.

Note that the projection $<k\mid z$ $ ; 2\beta >$ in \eqref{1.2} can be rewritten as
\begin{equation}
<k\mid z\frak{,}2\beta >=\left( K_{\beta }(z,z)  \right) ^{-\frac{%
1}{2}}h_{k}^{\beta }(z) ,  \label{1.3}
\end{equation}
where
\begin{equation}
h_{k}^{\beta }(z) :=\left( \frac{\Gamma (2\beta+k)
}{\pi \Gamma (2\beta) k!}\right) ^{\frac{1}{2}}z^{k},k=0,1,2,\cdots
\label{1.4}
\end{equation}
and
\begin{equation}
K_{\beta }(z,w) :=\pi ^{-1}\left( 1-z\overline{w}\right)
^{-2\beta },z,w\in \mathbb{D}  \label{1.5}
\end{equation}
are respectively an orthonormal basis and reproducing kernel of the weighted
Bergman space
\begin{equation}
\mathcal{A}_{\beta ,0}(\mathbb{D}) :=\left\{ \varphi \in
L^{2,\beta }(\mathbb{D}) ,\varphi \text{ holomorphic on }%
\mathbb{D}\right\} ,  \label{1.6}
\end{equation}
where $L^{2,\beta }(\mathbb{D}) $ denotes the Hilbert space of
functions $\varphi :$ $\mathbb{D\rightarrow C}$, which are square integrable
with respect to the measure $(1-|z|^{2}) ^{2\beta
-2}d\mu .$ Here $d\mu $ is the Lebesgue measure on  $\mathbb{D}.$

By another hand$,$ the Bergman space in \eqref{1.6} coincides with
the null space
\begin{equation}
\mathcal{A}_{\beta }(\mathbb{D}) =\left\{ \varphi \in L^{2,\beta
}(\mathbb{D}) ,\widetilde{H}_{\beta }\left[ \varphi \right]
=0\right\}  \label{1.7}
\end{equation}
of the second order differential operator
\begin{equation}
\widetilde{H}_{\beta }:=(1-|z|^{2}) ^{-\beta
}H_{\beta }(1-|z|^{2}) ^{\beta },  \label{1.8}
\end{equation}
where the \ operator $H_{\beta }$ is given by
\begin{equation}
\frac{1}{4}H_{\beta }:=-(1-|z|^{2}) ^{2}\frac{%
\partial ^{2}}{\partial z\partial \overline{z}}-\beta z\left( 1-\left|
z\right| ^{2}\right) \frac{\partial }{\partial z}+\beta \overline{z}\left(
1-|z|^{2}\right) \frac{\partial }{\partial \overline{z}}+\beta
^{2}|z|^{2}-\beta ^{2},  \label{1.9}
\end{equation}
and constitutes (in suitable units and up to additive constant) a
realization in $L^{2,0}(\mathbb{D}) $ of the Schr\"{o}dinger
operator with uniform magnetic field in $\mathbb{D}$, with a field strength
proportional to $\beta $ \cite{3}.

The spectrum of $\widetilde{H}_{\beta }$ in $L^{2,\beta }\left( \mathbb{D}%
\right) $ consists of eigenvalues of infinite multiplicity \textit{%
(hyperbolic Landau levels) }of the form:
\begin{equation}
\epsilon _{m}^{\beta }:=4(\beta-m) \left( 1-\beta +m\right) ,%
 m=0,1,2,\cdots ,\left[ \beta -\frac{1}{2}\right]   \label{1.10}
\end{equation}
provided that $2\beta >1.$ Here, $\left[ \eta \right] $ denotes the largest
integer not exceeding $\eta .$ As for the Bergman space $\mathcal{A}_{\beta
}(\mathbb{D}) $ in \eqref{1.7} the eigenspace
\begin{equation}
\mathcal{A}_{\beta ,m}(\mathbb{D}) :=\left\{ \varphi \in
L^{2,\beta }(\mathbb{D}) ,\widetilde{H}_{\beta }\left[ \varphi
\right] =\epsilon _{m}^{\beta }\varphi \right\}   \label{1.11}
\end{equation}
corresponding to the eigenvalue $\epsilon _{m}^{\beta }$ in $\left(
1.10\right) $ admits an orthogonal basis denoted $h_{k}^{\beta ,m}\left(
z\right) ,k=0,1,2,\cdots ,$ given in terms of Jacobi polynomials as well as a
reproducing kernel $K_{\beta ,m}(z,w) $ in an explicit form.
In this paper, we exploit these facts to construct a set of generalized
coherent states as
\begin{equation}
\mid z ;2\beta ,m>=\left( K_{\beta ,m}(z,z)  \right) ^{-%
\frac{1}{2}}\sum\limits_{k=0}^{+\infty }\frac{h_{k}^{\beta ,m}\left(
z\right) }{\sqrt{\rho _{\beta ,m}\left( k\right) }}\mid k>,z\in \mathbb{D}%
\text{,}  \label{1.12}
\end{equation}
where $\rho _{\beta ,m}\left( k\right) $ denotes the norm square of the
function $h_{k}^{\beta ,m}$ $(z) $ in $L^{2,\beta }\left( %
\mathbb{D}\right) .$ The states \eqref{1.12} enables us to attach
to each eigenspace $\mathcal{A}_{\beta ,m}(\mathbb{D}) $ a kind
of photon counting probability distribution in the same way as for the space
$\mathcal{A}_{\beta }(\mathbb{D}) =\mathcal{A}_{\beta ,0}\left( %
\mathbb{D}\right) $ corresponding to the lowest hyperbolic Landau level $m=0.
$ Indeed, for each fixed $m=0,1,2,\cdots ,\left[ \beta -\frac{1}{2}\right] $ and
$\lambda =|z|^{2},$ the probability mass function $P\left(
X=k\right) ,$ $k=0,1,2,\cdots $ of this counting random variable $X$ is obtained
as
\[
p_{k}\left( \lambda ,2\beta ;m\right) =\frac{\Gamma \left( 1+\frac{1}{2}%
( m+k-|m-k|) \right) \Gamma \left( 2\beta -m+\frac{1%
}{2}(|m-k| +k-m) \right) }{\Gamma \left( 1+\frac{1}{2%
}(m+k+|m-k|) \right) \Gamma \left( 2\beta -m-\frac{1%
}{2}(|m-k|+m-k) \right) }
\]
\begin{equation}
\times (1-\lambda) ^{2(\beta-m) }\lambda ^{\left|m-k\right| }\left( P_{\frac{1}{2}(m+k-|m-k|)
}^{\left( |m-k| ,2(\beta-m) -1\right) }(1-2\lambda) \right) ^{2}  \label{1.13}
\end{equation}
where $P_{\eta }^{\left( \tau ,\varsigma \right) }\left( .\right) $ denotes
the Jacobi polynomial \cite{4}. The probability distribution \eqref{1.13} can be considered as a kind of generalized negative
binomial probability distribution $X\sim \mathcal{NB}\left( \lambda ,2\beta
,m\right) $ depending on an additional parameter $m=0,1,2,\cdots ,\left[ \beta -%
\frac{1}{2}\right] .$ Thus, we study the main properties of the
family\thinspace of probability distributions in \eqref{1.13} and
we examine the quantum photon counting statistics with respect\thinspace \
to the location inside the disk $\mathbb{D}$ of the point labeling the
generalized coherent states $\mid z$ $;2\beta ,m>$ in \eqref{1.12}.

The paper is organized as follows. In Section 2, we recall briefly the
negative binomial states as well as their principal statistical properties.
Section 3 deals with some needed facts on the Shr\"{o}dinger operator with
magnetic field in the disk with an explicit description of some its needed
eigenspaces. In Section 4, we associate to each generalized Bergman space a
set of coherent states from which we obtain the announced probability
distribution. In section 5 \thinspace we give the main parameters of these
probability distributions and we discuss the classicality/nonclassicality of
the generalized coherent states with respect to the location of their
labeling points inside the disk.

\section{Negative binomial states}

The \textit{negative binomial states} are the field states that are
superposition of the number states with appropriately chosen coefficients.
They are intermediating states between a pure coherent state and a pure
thermal state and reduce to Susskind-Glogower phases states for a particular
limit\thinspace of the parameter \cite{4}. As mentioned above,
these states are labeled by points $z$ with $|z| <1$ and are of
the form
\begin{equation}
\mid z ;2\beta >=(1-|z|^{2}) ^{\beta
}\sum\limits_{k=0}^{+\infty }\left( \frac{\Gamma (2\beta+k) }{%
\Gamma (2\beta) k!}\right) ^{\frac{1}{2}}z^{k}\mid k>  \label{2.1}
\end{equation}
where $2\beta >1$ is a fixed parameter and $\mid k>$ are number states.The
states \eqref{2.1} are referred to as the negative binomial states
since their photon probability distribution:
\begin{equation}
\Pr \left( X=k\right) =(|z|^{2}) ^{k}\left(
1-|z|^{2}\right) ^{2\beta }\frac{\Gamma (2\beta+k)
}{\Gamma (2\beta) k!}  \label{2.2}
\end{equation}
obeys the negative binomial probability distribution, i.e., $X\sim \mathcal{%
NB}(\lambda ,2\beta) $ with parameters $\lambda =\left|
z\right| ^{2}$ and $2\beta >1.$ The mean number of photons and the variance
are given by $E(X) =\lambda 2\beta (1-\lambda) ^{-2}
$ and $Var(X) =\lambda 2\beta (1-\lambda) ^{-2}.$
The Mandel $Q$ parameter for the negative binomial states equals $\lambda
(1-\lambda) ^{-1}$ and is always positive since $0<\lambda <1.$
This means that photon statistics in the negative binomial states is always
super-Poissonian.

According to \cite{4}, we should mention some limiting
cases. For $\beta \rightarrow \infty ,|z| \rightarrow 0$ but $%
\beta |z| ^{-1}\rightarrow \mu $ \ the $\mathcal{NB}\left(
\lambda ,2\beta \right) $ reduces to the Poisson distribution $\mathcal{P}%
\left( \mu \right) $ characteristic of the coherent states of the harmonic
oscillator$.$ For $\beta \rightarrow 0,$ the photon number distribution $%
\mathcal{NB}(\lambda ,2\beta) $ reduces to the Bose-Einstein
distribution. When $|z| \rightarrow 0,$ $\mathcal{NB}\left(
\lambda ,2\beta \right) $ goes to Dirac's measure $\delta _{0}$ and the
negative binomial state in \eqref{2.1} goes to the vacuum state $%
\mid 0>.$

\section{An orthonormal basis in $\mathcal{A}_{\beta ,m}\left( \mathbb{D}%
\right) $.}

By $\left[ 3\right] $ the Schr\"{o}dinger operator on $\mathbb{D}$ with
constant magnetic field of strength proportional to $\beta >0$ can be
written as $:$%
\begin{equation}
\mathcal{L}_{\beta }:=-(1-|z|^{2}) ^{2}\frac{%
\partial ^{2}}{\partial z\partial \overline{z}}-\beta z\left( 1-\left|
z\right| ^{2}\right) \frac{\partial }{\partial z}+\beta \overline{z}\left(
1-|z|^{2}\right) \frac{\partial }{\partial \overline{z}}+\beta
^{2}|z|^{2}.  \label{3.1}
\end{equation}
which is also called Maass Laplacian on the disk. A slight modification of $%
\mathcal{L}_{B}$ is given by the operator
\begin{equation}
H_{\beta }:=4\mathcal{L}_{\beta }-4\beta ^{2}  \label{3.2}
\end{equation}
acting in the Hilbert space
\begin{equation}
L^{2,0}(\mathbb{D}) :=\left\{ \varphi :\mathbb{D\rightarrow C}%
,\int_{\mathbb{D}}\left| \varphi (z) \right| ^{2}\left( 1-\left|
z\right| ^{2}\right) ^{-2}d\mu (z) <+\infty \right\} ,  \label{3.3}
\end{equation}
The spectrum\ of $H_{\beta }$ in $L^{2,0}(\mathbb{D}) $ consists
of two parts: $(i)$\textit{\ }a continuous part $\left[ 1,+\infty \right[ $
, $(ii)$ a finite number of eigenvalues of the form ($\left[ 3\right] ):$%
\begin{equation}
\epsilon _{m}^{\beta }:=4(\beta-m) \left( 1-\beta +m\right) ,%
 m=0,1,2,\cdots ,\left[ \beta -\frac{1}{2}\right]  \label{3.4}
\end{equation}
with infinite degeneracy, provided that $2\beta >1.$ The eigenfunctions
corresponding to eigenvalues in \eqref{3.4} are known as bound
states. For our purpose, we shall consider the unitary equivalent
realization $\widetilde{H}_{\beta }$ of the operator $H_{\beta }$ in the
Hilbert space
\begin{equation}
L^{2,\beta }(\mathbb{D}) :=\left\{ \varphi :\mathbb{D\rightarrow
C},\int_{\mathbb{D}}\left| \varphi (z) \right| ^{2}\left(
1-|z|^{2}\right) ^{2\beta -2}d\mu (z) <+\infty
\right\} ,  \label{3.5}
\end{equation}
which is defined by
\begin{equation}
\mathsf{\ }\widetilde{H}_{\beta }:=\frak{T}_{\beta }^{-1}H_{\beta }\frak{T}%
_{\beta },  \label{3.6}
\end{equation}
where $\frak{T}_{\beta }:L^{2,\beta }(\mathbb{D}) \rightarrow
L^{2,0}(\mathbb{D}) $ is the unitary transformation defined by
the map $\varphi \mapsto (1-|z|^{2}) ^{-\beta
}\varphi .$

According to Eq. \eqref{3.6} the eigenspace of $H_{\beta }$ in $%
L^{2,0}(\mathbb{D}) $, which corresponds to the eigenvalue $%
\epsilon _{m}^{\beta }$ in \eqref{3.4}, is mapped isometrically via
the transform $\frak{T}_{\beta }$ onto the eigenspace
\begin{equation}
\mathcal{A}_{\beta ,m}(\mathbb{D}) :=\left\{ \Phi :\mathbb{D}%
\mathbf{\rightarrow }\mathbb{C},\Phi \in L^{2,\beta }\left( \mathbb{D}%
\right) \text{ and }\widetilde{H}_{\beta  }\Phi =\epsilon _{m}^{\beta
}\Phi \right\}  \label{3.7}
\end{equation}
These eigenspaces will play a central role in this work and some of their
spectral tools are summarized as follows:

\textbf{Proposition 3.1}. \textit{Let }$2\beta >1$\textit{\ and }$%
m=0,1,2,\cdots ,\left[ \beta -\frac{1}{2}\right] .$ \textit{Then, }

$\left( i\right) $ \textit{an orthogonal basis} of $\mathcal{A}_{\beta
,m}(\mathbb{D}) $\ \textit{is given by the set of functions}

\begin{equation}
\Phi _{k}^{\beta ,m}(z) :=|z| ^{|m-k|
}(1-|z|^{2}) ^{-m}e^{-i\left( m-k\right) \arg z}
\nonumber
\end{equation}
\begin{equation}
\times _{2}F_{1}\left( -m+\frac{m-k+|m-k| }{2},2\beta -m+\frac{%
|m-k| -m+k}{2},1+|m-k| ;|z|^{2}\right)
\label{3.8}
\end{equation}
$k=0,1,2,\cdots ,$\textit{where }$_{2}F_{1}\left( a,b,c;x\right) $\textit{\ is
the Gauss hypergeometric function } \cite{5}.

$\left( ii\right) $ \textit{the norm square }$\rho _{\beta ,m}\left(
k\right) $\textit{\ of the eigenfunction }$\Phi _{k}^{\beta ,m}$\textit{\ in
}$L^{2,\beta }(\mathbb{D}) $\textit{\ is given by}
\begin{equation}
\rho _{\beta ,m}\left( k\right) =\frac{\pi \left( \Gamma \left( 1+\left|
m-k\right| \right) \right) ^{2}}{\left( 2(\beta-m) -1\right) }%
\frac{\Gamma \left( m-\frac{|m-k| +m-k}{2}+1\right) \Gamma
\left( 2\beta -m-\frac{|m-k| +m-k}{2}\right) }{\Gamma \left( m+%
\frac{|m-k| -m+k}{2}+1\right) \Gamma \left( 2\beta -m+\frac{%
|m-k| -m+k}{2}\right) }.  \label{3.9}
\end{equation}
$\left( iii\right) $ \textit{the diagonal of the reproducing kernel\ of the
Hilbert\ }$\mathcal{A}_{\beta ,m}(\mathbb{D}) $\textit{\ is
given by }
\begin{equation}
K_{\beta ,m}(z,z)  =\pi ^{-1}\left( 2\beta -2m-1\right) \left(
1-|z|^{2}\right) ^{-2\beta },z\in \mathbb{D}\mathbf{.}
\label{3.10}
\end{equation}

\textbf{Proof.} \textit{\ }For\textit{\ }$\left( i\right) ,$ one can easily
chek that the functions $\Phi _{k}^{\beta ,m}(z) $ in \eqref{3.8} are of the form $\Phi _{k}^{\beta ,m}(z) =\frak{T}%
_{B}\left[ \phi _{k}^{\beta ,m}\right] (z) $, where $\phi
_{k}^{\beta ,m},k=0,1,2,\cdots ,$ is an orthonormal basis of the space
\begin{equation}
\mathcal{A}_{\beta ,m}^{0}(\mathbb{D}) :=\left\{ \phi :\mathbb{D}%
\mathbf{\rightarrow }\mathbb{C},\phi \in L^{2,0}(\mathbb{D})
\text{ and }H_{\beta  }\phi =\epsilon _{m}^{\beta }\phi \right\}
\label{3.11}
\end{equation}
as discussed in \cite[p. 9311]{9}, where the elements
of the basis have been labeled by an integer $j\geq -m$ and therefore one
has to take care of this by setting $k=j+m.$ By the fact that $\frak{T}%
_{\beta }$ is an isometry, on gets that $\Phi _{k}^{\beta ,m}(z)
,k=0,1,2,\cdots ,$ constitutes an orthonormal basis of $\mathcal{A}_{\beta
,m}(\mathbb{D}) .$ For $\left( ii\right) ,$ the square norm of
the eigenfunction $\phi _{k}^{\beta ,m}$ in the Hilbert space $L^{2,0}\left( %
\mathbb{D}\right) $ have been calculated in \cite[p.9313]{6} and remains the same for its image $\Phi _{k}^{\beta ,m}$
in $L^{2,\beta }(\mathbb{D}) $ under the unitary map $\frak{T}%
_{\beta }.$ For $\left( iii\right) $, it is not difficult to see that the
reproducing kernel $K_{\beta ,m}(z,w) $ of the Hilbert space $%
\mathcal{A}_{\beta ,m}(\mathbb{D}) $ reads
\begin{equation}
K_{\beta ,m}(z,w) =(1-|z|^{2}) ^{-\beta}K_{\beta ,m}^{0}(z,w) (1-|w|^{2})^{-\beta }  \label{3.12}
\end{equation}
where $K_{\beta ,m}^{0}(z,w) $ denotes the reproducing kernel of
the Hilbert space $\mathcal{A}_{\beta ,m}^{0}(\mathbb{D}) $ in
\eqref{3.11}, whose diagonal term is given by the function \cite[p.9313]{6}:
\begin{equation}
K_{\beta ,m}^{0}(z,z)  =\pi ^{-1}\left( 2\beta -2m-1\right) ,z\in %
\mathbb{D}.  \label{3.13}
\end{equation}
The proof of proposition is finished.$\Box $

\textit{\ }

We should note that in the case $m=0,$ the eigenspace $\mathcal{A}_{\beta
,0}(\mathbb{D}) $ coincides with the weighted Bergmann space on
the disk defined in \eqref{1.6}. \ Being motivated by this remark,
the eigenspace $\mathcal{A}_{\beta ,m}(\mathbb{D}) $ of $%
\widetilde{H}_{\beta }$ \ corresponding to the eigenvalue $\epsilon
_{m}^{\beta }$\ given\textit{\ }in\textit{\ }(3.2) will be called\textit{\
generalized weighted Bergman space of index }$m.$

\section{Coherent states and probability distributions}

In this section, we present a generalization of coherent states according to
the procedure in \cite{7}. For this, let $(X,\sigma )$\ be a
measure space and let $\mathcal{A}^{2}\subset L^{2}(X,\sigma )$\ be a closed
subspace of infinite dimension. Let $\left\{ f_{n}\right\} _{n=0}^{\infty }$
be an orthogonal basis of $\mathcal{A}^{2}$ satisfying, for arbitrary $u\in
X,$

\begin{equation}
\omega \left( u\right) :=\sum_{n=0}^{\infty }\rho _{n}^{-1}\left|
f_{n}\left( u\right) \right| ^{2}<+\infty ,  \label{4.1}
\end{equation}
where $\rho _{n}:=\left\| f_{n}\right\| _{L^{2}(X,\sigma )}^{2}$. Define

\begin{equation}
\frak{K}(u,v):=\sum_{n=0}^{\infty }\frac{1}{\rho _{n}}f_{n}\left( u\right)
\overline{f_{n}(v)},~u,v\in X.  \label{4.2}
\end{equation}
Then, $\frak{K}(u,v)$\ is a reproducing kernel, $\mathcal{A}^{2}$ is the
corresponding reproducing kernel Hilbert space and $\omega \left( u\right) :=%
\frak{K}(u,u)$, $u\in X$.

\textbf{Definition. 4.1}. \textit{Let }$\mathcal{H}$\textit{\ be a Hilbert
space with }$\dim \mathcal{H}=\infty $\textit{\ and }$\left\{ \phi
_{n}\right\} _{n=0}^{\infty }$\textit{\ be an orthonormal basis of }$%
\mathcal{H}.$\textit{\ The coherent states labeled by points }$u\in X$%
\textit{\ are defined as the ket-vectors }$\vartheta _{u}\equiv \mid u>\in
\mathcal{H}:$%
\begin{equation}
\vartheta _{u}\equiv \mid u>:=\left( \omega \left( u\right) \right) ^{-\frac{%
1}{2}}\sum_{n=0}^{\infty }\frac{f_{n}\left( u\right) }{\sqrt{\rho _{n}}}\phi
_{n}.\quad \quad   \label{4.3}
\end{equation}
Now, by Definition 4.1, it is straightforward to show that $%
<u\mid u>=1$\ and the coherent state transform $W:\mathcal{H\rightarrow A}%
^{2}\subset L^{2}(X,\sigma )$ defined by
\begin{equation}
W\left[ \phi \right] \left( u\right) :=\left( \omega \left( u\right) \right)
^{\frac{1}{2}}<\vartheta _{u}\mid \phi >\quad \quad   \label{4.4}
\end{equation}
is an isometry. Thus, for $\phi ,\psi \in \mathcal{H}$, we have
\[
<\phi \mid \psi >_{\mathcal{H}}=<W\left[ \phi \right] \mid W\left[ \psi
\right] >_{L^{2}\left( X,\sigma \right) }=\int\limits_{X}d\sigma \left(
u\right) \omega \left( u\right) <\phi \mid \vartheta _{u}><\vartheta
_{u}\mid \psi >.
\]
Thereby, we have a resolution of the identity of $\mathcal{H}$ which can be
expressed in Dirac's bra-ket notation as
\begin{equation}
\mathbf{1}_{\mathcal{H}}=\int\limits_{X}d\sigma \left( u\right) \omega
\left( u\right) \mid u><u\mid ,\quad \quad   \label{4.5}
\end{equation}
and where $\omega \left( u\right) $\ appears as a weight function.

Now, we are in position to construct for each hyperbolic Landau level $%
\epsilon _{m}^{B}$ given in \eqref{3.4} a set of generalized coherent states
according to formula \eqref{4.3} as
\begin{equation}
\mid z,2\beta ,m>:=\left( K_{\beta ,m}(z,z)  \right) ^{-\frac{1}{2%
}}\sum_{k=0}^{+\infty }\frac{\Phi _{k}^{\beta ,m}(z) }{\sqrt{%
\rho _{\beta ,m}\left( k\right) }}\mid k,\alpha >  \label{4.6}
\end{equation}
with the following meaning:

\begin{quote}
$\cdot $  $(X,\sigma )=(\mathbb{D},\left( 1-|z|
^{2}\right) ^{2\beta -2}d\mu (z) ),$ $d\sigma (z)
=(1-|z|^{2}) ^{2\beta -2}d\mu (z) ,$ $%
d\mu (z) $ being the Lebesgue measure on $\mathbb{D},$

$\mathbf{\cdot }$ $\mathcal{A}^{2}:=\mathcal{A}_{\beta ,m}\left( \mathbb{D}%
\right) $ \ denotes the eigenspace of $\widetilde{H}_{\beta  }$in $%
L^{2,\beta }(\mathbb{D}) $,

$\cdot $ $K_{\beta ,m}(z,z)  $  $=\pi ^{-1}\left( 2\beta
-2m-1\right) (1-|z|^{2}) ^{-2\beta },$

$\cdot $  $\Phi _{k}^{\beta ,m}(z)$ are the eigenfunctions given by
\eqref{3.8} in terms of the Gauss hypergeometric function $%
_{2}F_{1}\left( .\right) $

$\cdot $  $\rho _{\beta ,m}\left( k\right) $\ being the norm square
of $\Phi _{k}^{\beta ,m}$ given in \eqref{3.9},

$\mathbf{\cdot }$ $\mathcal{H}$:=$L^{2}(\mathbb{R}_{+}^{\ast },x^{-1}dx)$ is
the Hilbert space carrying the coherent states \eqref{5.6},

$\cdot $  $\mid k,\alpha >\equiv \psi _{k}^{\alpha },k=0,1,2,\cdots ,$%
is the complete orthonormal basis of $L^{2}(\mathbb{R}_{+}^{*},x^{-1}dx)$
consisting of functions given by \cite{8}:
\[
\psi _{k}^{\alpha }\left( x\right) :=\left( \frac{\Gamma \left( k+2\alpha
\right) }{k!}\right) ^{-\frac{1}{2}}x^{\alpha }\exp \left( -\frac{1}{2}%
x\right) L_{k}^{\left( 2\alpha -1\right) }\left( x\right) ,x\in \mathbb{R}%
_{+}^{\ast }
\]
where $L_{k}^{\left( \eta \right) }\left( .\right) $ denotes the generalized
Laguerre polynomial \cite{5}.
\end{quote}

\textbf{Definition 4.2}. \textit{For each fixed }$m=0,1,2,\cdots ,\left[ \beta -%
\frac{1}{2}\right] .$\textit{\ The coherent states }$\left( \mid z,\beta
,m>\right) _{z\in \mathbb{D}}$ \textit{associated with the generalized
Bergman space }$\mathcal{A}_{B,m}(\mathbb{D}) $\textit{\ are
defined as a superposition of the basis }$\psi _{k}^{\alpha }$ of the Hilbert
space $L^{2}(\mathbb{R}_{+}^{*},x^{-1}dx)$ \textit{through the wave
functions}

\[
<x\mid z,2\beta ,m>:=\frac{\sqrt{\pi }}{\sqrt{2\beta -2m-1}}\left( 1-\left|
z\right| ^{2}\right) ^{\beta -m}
\]

\begin{equation}
\times \sum_{k=0}^{+\infty }\frac{|z| ^{|m-k|
}e^{-i\left( m-k\right) \arg z}}{\sqrt{\rho _{\beta ,m}\left( k\right) }}%
P_{\min \left( m,k\right) }^{\left( |m-k| ,2\left( \beta
-m\right) -1\right) }\left( 1-2|z|^{2}\right) \psi _{k}^{\alpha
}\left( x\right) ,x\in \mathbb{R}_{+}^{*}.  \label{4.7}
\end{equation}

Now, in view of \eqref{4.7} the projection of the coherent states $%
\mid z,\beta ,m>$ onto the state $\psi _{k}^{\alpha }$ is given by the
scalar product
\begin{equation}
\left\langle z,2\beta ,m\mid \psi _{k}^{\alpha }\right\rangle _{\mathcal{H}%
}=\left( K_{\beta ,m}(z,z)  \right) ^{-\frac{1}{2}}\frac{\Phi
_{k}^{\beta ,m}(z) }{\sqrt{\rho _{k}^{\beta ,m}\ }},k=0,1,2,\cdots .
\label{4.8}
\end{equation}
Therefore, the squared modulus of $\left\langle z,2\beta ,m\mid \psi
_{k}^{\alpha }\right\rangle _{\mathcal{H}}$ gives the probability that $k$
photons will be found in the coherent state $\mid z,2\beta ,m>$ .This leads
to the mass distribution
\begin{equation}
p_{k}\left( |z|^{2},2\beta ,m\right) :=\left| \left\langle
z,2\beta ,m\mid \psi _{k}^{\alpha }\right\rangle _{\mathcal{H}}\right|
^{2},\quad ~k=0,1,2,\ldots ,  \label{4.9}
\end{equation}
which is denoted $p_{k}\left( \lambda ,2\beta ,m\right) $ with $\lambda
=|z|^{2}.$ Being motivated by this quantum probability, we then
write:

\textbf{Definition 4.3}. \textit{For each fixed }$m=0,1,2,\cdots ,\left[ \beta -%
\frac{1}{2}\right] $ \textit{the discrete random variable }$X$\textit{\ with
the probability distribution }
\[
p_{k}\left( \lambda ,2\beta ;m\right) :=\gamma _{\beta ,m,k}\left( 1-\lambda
\right) ^{2(\beta-m) }\lambda ^{|m-k| }\left( P_{%
\frac{1}{2}\left( m+k-|m-k| \right) }^{\left( |m-k|
,2(\beta-m) -1\right) }\left( 1-2\lambda \right) \right) ^{2}
\]
\textit{with}
\begin{equation}
\gamma _{\beta ,m,k}:=\frac{\Gamma \left( 1+\frac{1}{2}\left( m+k-\left|
m-k\right| \right) \right) \Gamma \left( 2\beta -m+\frac{1}{2}\left( \left|
m-k\right| +k-m\right) \right) }{\Gamma \left( 1+\frac{1}{2}\left(
m+k+|m-k| \right) \right) \Gamma \left( 2\beta -m-\frac{1}{2}%
(|m-k|+m-k) \right) }  \label{4.10}
\end{equation}
\textit{and denoted by }$X\sim \mathcal{NB}\left( \lambda ,2\beta ;m\right)
,~\lambda >0$ and $2\beta >1$\textit{\ will be called the extended negative
binomial probability distribution\ attached to the generalized Bergman space
}$\mathcal{A}_{\beta ,m}(\mathbb{D}) .$

\textbf{Remark 4.1. \ }Note that for $m=0,$ the above expression in $\left(
5.10\right) $ reduces to
\begin{equation}
p_{k}\left( \lambda ,2\beta ;0\right) =(1-\lambda) ^{2\beta
}\lambda ^{k}\frac{\Gamma (2\beta+k) }{k!\Gamma \left( 2\beta
\right) },k=0,1,2,\cdots   \label{4.11}
\end{equation}
which is the standard negative binomial distribution $\mathcal{NB}\left(
\lambda ,2B\right) $ with parameter $\lambda $ and $2\beta $ in \eqref{2.1}

\textbf{Remark 4.2. }We should note that expression of the mass distribution
$p_{k}\left( \lambda ,2\beta ,m\right) $ in $\left( 4.10\right) $ may also
appear when calculating Franck-Condon factors in special case of molecular
vibration described by the Morse potential \cite[p.6]{9}.

\section{The generating function of $X\sim \mathcal{NB}\left( \lambda
,2\beta ;m\right) $ and photon number statistics}

The purpose of this section is to give some essential parameters of \textit{%
\ }$X\sim \mathcal{NB}\left( \lambda ,2\beta ;m\right) $ . We first
determine the generating function

\begin{equation}
G_{X}^{m}(\xi )=\sum_{k=0}^{+\infty }\xi ^{k}p_{k}\left( \lambda ,2\beta
;m\right)   \label{5.1}
\end{equation}
as a convenient way to obtain information about this random variable.

\textbf{Proposition 6.1.} \textit{Let }$m=0,1,2,\cdots ,\left[ \beta -\frac{1}{2}%
\right] .$ \textit{Then the generating function of the random variable }$%
X\sim \mathcal{NB}\left( \lambda ,2\beta ;m\right) $\textit{\ \ is given by}
\begin{equation}
G_{X}^{m}(\xi )=\left( \frac{1-\lambda }{1-\lambda \xi }\right) ^{2\beta
}\left( \frac{\left( \lambda -\xi \right) \left( 1-\lambda \xi \right) }{%
(1-\lambda) ^{2}}\right) ^{m}P_{m}^{\left( 2\left( \beta
-m\right) -1,0\right) }\left( 1+\frac{2\xi (1-\lambda) ^{2}}{%
\left( \lambda -\xi \right) \left( 1-\lambda \xi \right) }\right)   \label{5.2}
\end{equation}

\textbf{Proof.} \ The integer $m=0,1,\cdots ,\left[ \beta -\frac{1}{2}\right] $
being fixed, we start by writing the generating function of $X\sim \mathcal{%
NB}\left( \lambda ,2\beta ;m\right) $ according to $\left( 5.1\right) $ and
we make use of definition $\left( 4.3\right) $, we have that
\begin{equation}
G_{X}^{m}(\xi )=\sum_{k=0}^{+\infty }\gamma _{\beta ,m,k}\xi ^{k}\left(
1-\lambda \right) ^{2(\beta-m) }\lambda ^{|m-k|
}\left( P_{\frac{1}{2}\left( m+k-|m-k| \right) }^{\left( \left|
m-k\right| ,2(\beta-m) -1\right) }\left( 1-2\lambda \right)
\right) ^{2}  \label{5.3}
\end{equation}
We split this sum into two part as
\begin{equation}
G_{X}^{m}(\xi )=\mathcal{G}_{\beta ,m,\lambda }^{(<\infty )}\left( \xi
\right) +\mathcal{G}_{\beta ,m,\lambda }^{\left( \infty \right) }\left( \xi
\right)   \label{5.4}
\end{equation}
where $\mathcal{G}_{\beta ,m,\lambda }^{(<\infty )}\left( \xi \right) $
denotes \ a finite sum given by
\begin{eqnarray*}
\mathcal{G}_{\beta ,m,\lambda }^{(<\infty )}\left( \xi \right)
&:&=\sum_{j=0}^{m-1}(1-\lambda) ^{2(\beta-m) }\xi
^{k}(\frac{k!}{m!}\frac{\Gamma \left( 2\beta -m\right) }{\Gamma \left(
2\beta -2m+k\right) }\lambda ^{m-k}\left( \left( P_{k}^{\left( m-k,2\left(
\beta -m\right) -1\right) }\left( 1-2\lambda \right) \right) ^{2}\right)  \\
&&-\frac{m!}{k!}\frac{\Gamma \left( 2\beta -2m+k\right) }{\Gamma \left(
2\beta -m\right) }\lambda ^{k-m}\left( \left( P_{m}^{\left( k-m,2\left(
\beta -m\right) -1\right) }\left( 1-2\lambda \right) \right) ^{2}\right)
\end{eqnarray*}
and $\mathcal{G}_{\beta ,m,\lambda }^{\left( \infty \right) }\left( \xi
\right) $ denotes the following infinite sum:
\begin{equation}
\mathcal{G}_{\beta ,m,\lambda }^{\left( \infty \right) }\left( \xi \right)
:=\sum_{k=0}^{+\infty }\xi ^{k}\frac{m!}{k!}\frac{\Gamma \left( 2\beta
-2m+k\right) }{\Gamma \left( 2\beta -m\right) }(1-\lambda)
^{2(\beta-m) }\lambda ^{k-m}\left( \left( P_{m}^{\left(
k-m,2(\beta-m) -1\right) }\left( 1-2\lambda \right) \right)
^{2}\right)   \label{5.5}
\end{equation}
Noting that the finite sum $\mathcal{G}_{\beta ,m,\lambda }^{\left( <\infty
\right) }\left( \xi \right) $ contains the following difference
\begin{equation}
\frac{\lambda ^{m-k}\left( P_{k}^{\left( m-k,2(\beta-m)
-1\right) }\left( 1-2\lambda \right) \right) ^{2}}{\left( k!\Gamma \left(
2\beta -m\right) \right) ^{-1}m!\Gamma \left( 2\beta -2m+k\right) }-\frac{%
\lambda ^{k-m}\left( P_{m}^{\left( k-m,2(\beta-m) -1\right)
}\left( 1-2\lambda \right) \right) ^{2}}{\left( m!\Gamma \left( 2\beta
-2m+k\right) \right) ^{-1}k!\Gamma \left( 2\beta -m\right) }.  \label{5.6}
\end{equation}
The latter suggests to make use of the identity $\left( \left[ 10\right] ,%
 p.63\right) $:
\begin{equation}
\frac{\Gamma \left( n+1\right) }{\Gamma \left( n-l+1\right) l!}P_{n}^{\left(
-l,\nu \right) }\left( u\right) =\frac{\Gamma \left( n+\nu +1\right) }{%
l!\Gamma \left( n+\nu -l+1\right) }\left( \frac{u-1}{2}\right)
^{l}P_{n-l}^{\left( l,\nu \right) }\left( u\right) ,~1\leq l\leq n  \label{5.7}
\end{equation}
for $k=n$ , $l=k-m$, \ $u=1-2\lambda $ and $\nu =2(\beta-m) -1.$
We then write
\begin{equation}
P_{k}^{\left( m-k,\nu \right) }\left( 1-2\lambda \right) =\frac{m!\Gamma
\left( 2\beta -2m+k\right) }{\left( -1\right) ^{m-k}k!\Gamma \left( 2\beta
-m\right) \lambda ^{m-k}}P_{m}^{\left( k-m,\nu \right) }\left( 1-2\lambda
\right) .  \label{5.8}
\end{equation}
After calculation, we obtain that $\mathcal{G}_{\beta ,m,\lambda }^{(<\infty
)}\left( \xi \right) $ $=0$. Therefore, it remains to calculate the infinite
sum which reads
\begin{equation}
\mathcal{G}_{\beta ,m,\lambda }^{(<\infty )}\left( \xi \right) =\Upsilon
_{\beta ,m}^{1}\left( \lambda ,\xi \right) \sum_{k=0}^{+\infty }\frac{\Gamma
\left( 2\beta -2m+k\right) }{k!}(\xi \lambda )^{k-m}\left( P_{m}^{\left(
k-m,\nu \right) }\left( 1-2\lambda \right) \right) ^{2}  \label{5.9}
\end{equation}
where the prefactor is given by
\begin{equation}
\Upsilon _{\beta ,m}^{1}\left( \lambda ,\xi \right) :=\frac{m!\left(
1-\lambda \right) ^{2(\beta-m) }\xi ^{m}}{\Gamma \left( 2\beta
-m\right) }  \label{5.10}
\end{equation}
If we put $\tau =\xi \lambda $ and $k-m=s,$ we will need to calculate the
sum
\begin{equation}
\mathcal{S}:=\sum_{s\geq -m}\frac{\Gamma \left( 2\beta -m+s\right) }{\left(
s+m\right) !}\tau ^{s}\left( P_{m}^{\left( s,\nu \right) }\left( u\right)
\right) ^{2}  \label{5.11}
\end{equation}
where $u=1-2\lambda $ and $\nu =2(\beta-m) -1.$ Once again, we
make use of the identity $\left( 5.7\right) $ to rewrite the sum $\left(
5.11\right) $ as follows
\begin{equation}
\mathcal{S}=\Upsilon _{\beta ,m}^{2}\left( u,\tau \right)
\sum\limits_{j=0}^{+\infty }\frac{j!}{\left( \nu +1\right) _{j}}\left(
P_{j}^{\left( m-j,\nu \right) }\left( u\right) \right) ^{2}\left( \frac{%
4\tau }{\left( u-1\right) ^{2}}\right) ^{j}  \label{5.12}
\end{equation}
where the prefactor
\begin{equation}
\Upsilon _{\beta ,m}^{2}\left( u,\tau \right) :=\left( \frac{\Gamma \left(
2\beta -m\right) }{m!}\right) ^{2}\tau ^{-m}\left( \frac{u-1}{2}\right) ^{2m}
\label{5.13}
\end{equation}
Making use of the following identity due to Srivastava and Rao \cite[p. 1329]{11}:
\[
\sum_{n=0}^{+\infty }\frac{n!t^{n}}{\left( 1+\beta _{0}\right) _{n}}%
P_{m}^{(\gamma -n,\beta _{0})}\left( x\right) P_{m}^{(\gamma -n,\beta
_{0})}\left( y\right) =\frac{\left( 1-t\right) ^{\gamma }}{\left( 1-\frac{%
\left( x-1\right) \left( y-1\right) t}{4}\right) ^{1+\gamma +\beta _{0}}}
\]
\begin{equation}
\times _{2}\digamma _{1}\left( 1+\gamma +\beta _{0},-\gamma ,1+\beta _{0};%
\frac{-\left( x+1\right) \left( y+1\right) t}{\left( 1-t\right) \left(
4-\left( x-1\right) \left( y-1\right) t\right) }\right)   \label{5.14}
\end{equation}
for $t=\frac{4\tau }{\left( u-1\right) ^{2}}$, $x=y=u$ , $\beta _{0}=\nu
,\gamma =m$ and $n=j,$ we obtain after computation and summarizing up the
above steps
\begin{eqnarray}
G_{X}^{m}\left( \xi \right)  &=&\frac{\Gamma \left( 2\beta -m\right) }{%
m!\Gamma \left( 2(\beta-m) \right) }\left( \frac{1-\lambda }{%
1-\lambda \xi }\right) ^{2\beta }\left( \frac{\left( \lambda -\xi \right)
\left( 1-\lambda \xi \right) }{(1-\lambda) ^{2}}\right) ^{m}
\label{5.15} \\
&&\times _{2}\digamma _{1}\left( -m,2\beta -m,2(\beta-m) ;\frac{%
-\xi (1-\lambda) ^{2}}{\left( \lambda -\xi \right) \left(
1-\lambda \xi \right) }\right)   \nonumber
\end{eqnarray}
Finally, by the help of the relation \cite{5}:
\[
_{2}F_{1}\left( k+\nu +\varrho +1,-k,1+\nu ;\frac{1-t}{2}\right) =\frac{%
k!\Gamma \left( 1+\nu \right) }{\Gamma \left( k+1+\nu \right) }P_{k}^{(\nu
,\varrho )}\left( t\right)
\]
connecting the hypergeometric function $_{2}F_{1}$ $\left( .\right) $ with
the Jacobi polynomial $P_{k}^{(\nu ,\varrho )}\left( .\right) $, we arrive
at the announced expression of the generating function $G_{X}^{m}\left( \xi
\right) .$ This ends the proof of Proposition 5.1. $\Box $

\textbf{Remark 5.1.\ }Note that for $m=0,$ the expression in \eqref{6.2} reduces to
\begin{equation}
G_{X}^{0}\left( \xi \right) =\left( \frac{1-\lambda \xi }{1-\lambda }\right)
^{-2\beta }  \label{5.16}
\end{equation}
which is the well known characteristic function of standard negative
binomial distribution $\mathcal{NB}(\lambda ,2\beta) $with
parameter $\lambda $ and $2\beta .$

\textbf{Corollary 5.1.} \ \textit{Let }$m=0,1,2,\cdots \left[ \beta -\frac{1}{2}%
\right] $\textit{\ Then the mean value and the variance of the random
variable }$X\sim \mathcal{NB}\left( \lambda ,2\beta ;m\right) $\textit{\ are
respectively given by }
\begin{eqnarray}
E(X)  &=&\frac{2\lambda \beta }{1-\lambda }+m  \label{5.17} \\
Var(X)  &=&\frac{2\lambda }{(1-\lambda) ^{2}}+\frac{%
m}{(1-\lambda) ^{2}}\lambda \left( \beta -2-\frac{\lambda }{2}%
\right)   \nonumber
\end{eqnarray}
\textbf{Proof.} We make use of the expression of \ the generating function $%
G_{X}^{m}\left( \xi \right) $ obtained in Proposition 6.1 to derive the mean value
the mean value of the random variable $X\sim \mathcal{NB}\left( \lambda
,2\beta ;m\right) $ through the relation
\begin{equation}
E(X) =\frac{\partial }{\partial \xi }\left(
G_{X}^{m}\left( \xi \right) \right) \mid _{\xi =1}{}{}{}  \label{5.18}
\end{equation}
Straightforward calculations give
\begin{equation}
\frac{\partial }{\partial \xi }\left( G_{X}^{m}\left( \xi \right) \right)
\mid _{\xi =1}=\frac{2\lambda \beta }{1-\lambda }+m.  \label{5.19}
\end{equation}
The variance is also obtained by using the well known fact that
\begin{equation}
Var(X) =E\left( X^{2}\right) -\left( E(X) \right)
^{2}  \label{5.20}
\end{equation}
where $E\left( X^{2}\right) $ can also be obtained from the the generating
function $G_{X}^{m}\left( \xi \right) $ as
\begin{equation}
E\left( X^{2}\right) ={}{}{}\frac{\partial }{\partial \xi }\left(
G_{X}^{m}\left( \xi \right) \right) \mid _{\xi =1}+\frac{\partial ^{2}}{%
\partial \xi ^{2}}\left( G_{X}^{m}\left( \xi \right) \right) \mid _{\xi =1}
\label{5.21}
\end{equation}
After tedious calculations we obtain that
\begin{equation}
E\left( X^{2}\right) =\frac{2\beta \left( 2\beta +1\right) \lambda
^{2}+4m(1-\lambda) \lambda \beta }{(1-\lambda) ^{2}%
}+m\left( m-1-\frac{2\lambda }{(1-\lambda) ^{2}}\right)
\label{5.22}
\end{equation}
\[
+\frac{m\lambda \left( 2\beta -\lambda \right) }{2(1-\lambda)
^{2}}.
\]
Substituting \eqref{5.22} and \eqref{5.19} in \eqref{5.20}, we arrive at
\[
Var(X) =\frac{2\lambda \beta }{(1-\lambda) ^{2}}+%
\frac{m}{(1-\lambda) ^{2}}\lambda \left( \beta -2-\frac{\lambda
}{2}\right)
\]
This ends the proof of the corollary.$\Box $

\textbf{Remark 5.2. }For $m=0,$ the result of corollary 6.1 reads $E\left(
X\right) =2\beta \lambda (1-\lambda) ^{-2}$ and $Var\left(
X\right) =2\beta \lambda (1-\lambda) ^{-2}$ which are known
parameters of the standard negative binomial probability distribution.

\section{Photon counting statistics}

To define a measure of non classicality \ of a quantum states one can follow
several different approach$.$ An earlier attempt to shed some light on the
non-classicality of a quantum state was pioneered by Mandel \cite{12}, who investigated radiation fields and introduced the parameter
\begin{equation}
Q=\frac{Var(X) }{E(X) }-1,  \label{6.1}
\end{equation}
to measure the deviation of the photon number statistics from the Poisson
distribution , characteristic of coherent states. Indeed, $Q=0$ characterize
Poissonian statistics. If $Q<0$ we have \textit{sub-Poissonian} statistics
otherwise, statistics are \textit{super-Poissonian. }

In our context, as mentioned in section 1, if $m=0$ then \textit{\ }$X\sim
\mathcal{NB}(\lambda ,2\beta) $ obeys the negative binomial
distribution and the corresponding photon counting statistics are
super-Poissonian. For $m\neq 0$ we make use of the statistical parameters
obtained in corollary 5.1 to calculate Mandel parameter $Q(X) $
corresponding the random variable \textit{\ }$X\sim \mathcal{NB}\left(
\lambda ,2\beta ;m\right) $ and we summarize the discussion with respect to
the sign of $Q(X) $ in the following statement:

\textbf{Proposition 6.1}. \textit{Let }$m=1,2,\cdots ,\left[ \beta -\frac{1}{2}%
\right] .$\textit{\ Then, the photon counting statistics are :}

$\left( i\right) $\textit{\ sub-Poissonian for points }$z\in \frak{D}$%
\textit{\ belonging to the open disk }$\frak{D}\left( 0,r_{\beta ,m}\right) $%
\textit{\ or radius }
\begin{equation}
r_{\beta ,m}:=\frac{-m\beta +\sqrt{\left( \beta ^{2}-6\right) m^{2}+8\beta m}%
}{4\beta -3m}  \label{6.2}
\end{equation}
\textit{For such labeling point }$z$\textit{\ the states }$\mid z,\beta ,m>$%
\textit{\ are non-classical.}

$\left( ii\right) $ \textit{Poissonian for points }$z$\textit{\ of the\ the
boundary disk }$\partial \frak{D}\left( 0,r_{\beta ,m}\right) .$\textit{\
Here the states }$\mid z,2\beta ,m>$\textit{\ becomes pure coherent states.}

$\left( iii\right) $ \textit{Super-Poissonian for }$z\in \mathbb{D}\setminus
\frak{D}\left( 0,r_{\beta ,m}\right) .$\textit{\ For such points the states }%
$\mid z,2\beta ,m>$\textit{\ may describe thermal (or chaotic) light.}

\textbf{Proof.} Making use of corollary 6.1, the Mandel parameter $\left(
6.1\right) $ corresponding to the random variable $X\sim \mathcal{NB}\left(
\lambda ,2\beta ;m\right) $ has the following expression
\begin{equation}
Q(X) =\frac{\left( 4\beta -3m\right) \lambda ^{2}+2\beta \lambda
m-2m}{2(1-\lambda) \left( 2\beta \lambda -m\lambda +m\right) }.
\label{6.3}
\end{equation}
We look at the roots of the equation
\begin{equation}
\left( 4\beta -3m\right) \lambda ^{2}+2\beta \lambda m-2m=0  \label{6.4}
\end{equation}
with respect to the variable $\lambda $ with $0<\lambda <1.$ The
discriminant $\Delta ^{\prime }=\beta ^{2}m^{2}+2m\left( 4\beta -3m\right) >0
$ since $0$ $\leq m\leq \left[ \beta -\frac{1}{2}\right] $ and one can
easily see that roots of  Eq. \eqref{6.4} are of the form:
\begin{equation}
\lambda _{\pm }\left( \beta ,m\right) :=\frac{-mB\pm \sqrt{\left( \beta
^{2}-6\right) m^{2}+8\beta m}}{4\beta -3m}  \label{7.5}
\end{equation}
But the only \textit{admissible} root in the sense that it belongs to the
interval $\left] 0,1\right[ $ is $\lambda _{+}.$ We put $r_{\beta
,m}:=\lambda _{+}\left( \beta ,m\right) $ and the assertions $\left(
i\right) ,\left( ii\right) $ and $\left( iii\right) $ follow by discussing
the sign of the parameter $Q(X) $ in \eqref{6.3}. $\Box $

\end{document}